\documentclass[draftcls,onecolumn,12pt]{IEEEtran}

\usepackage{epsfig}
\usepackage{amsmath}
\usepackage{amsfonts}
\usepackage{amssymb}
\usepackage{subfigure}

\newcommand{\Pout}{P_{\rm out}(\snr,R)}
\newcommand{\snr}{{\sf SNR}}
\newcommand{\diag}{{\hbox{diag}}}
\newcommand{\openone}{\leavevmode\hbox{\small1\normalsize\kern-.33em1}}

\newcommand{\beq}{\begin{equation}}
\newcommand{\eeq}{\end{equation}}
\newcommand{\eqdef}{\stackrel{\Delta}{=}}

\newcommand{\sv}{{\boldsymbol s}}
\newcommand{\xv}{{\boldsymbol x}}
\newcommand{\yv}{{\boldsymbol y}}
\newcommand{\zv}{{\boldsymbol z}}

\newcommand{\alphav}{{\boldsymbol \alpha}}
\newcommand{\deltav}{{\boldsymbol \delta}}
\newcommand{\epsilonv}{{\boldsymbol \epsilon}}

\newcommand{\zerov}{{\boldsymbol 0}}
\newcommand{\onev}{{\boldsymbol 1}}

\newcommand{\Hm}{{\boldsymbol H}}
\newcommand{\Id}{{\boldsymbol I}}
\newcommand{\Mm}{{\boldsymbol M}}
\newcommand{\Sm}{{\boldsymbol S}}
\newcommand{\Xm}{{\boldsymbol X}}
\newcommand{\Ym}{{\boldsymbol Y}}
\newcommand{\Zm}{{\boldsymbol Z}}

\newcommand{\CC}{\mathbb{C}}
\newcommand{\EE}{\mathbb{E}}
\newcommand{\FF}{\mathbb{F}}
\newcommand{\RR}{\mathbb{R}}
\newcommand{\ZZ}{\mathbb{Z}}

\newcommand{\Cc}{{\cal C}}
\newcommand{\Ec}{{\cal E}}
\newcommand{\Kc}{{\cal K}}
\newcommand{\Nc}{{\cal N}}
\newcommand{\Oc}{{\cal O}}
\newcommand{\Sc}{{\cal S}}
\newcommand{\Xc}{{\cal X}}

\newtheorem{definition}{Definition}
\newtheorem{theorem}{Theorem}
\newtheorem{proposition}{Proposition}

\newtheorem{lemma}{Lemma}

\title{Multidimensional Coded Modulation in Block-Fading Channels} 
\author{Albert Guill\'en i F\`abregas
\thanks{A. Guill\'en i F\`abregas is with the Department of Engineering, University of Cambridge, Trumpington Street, Cambridge CB2 1PZ, UK, e-mail: {\tt guillen@ieee.org}.} and Giuseppe Caire
\thanks{G. Caire is with the Electrical Engineering Department, University of Southern California, 3740 McClintock Ave., Los Angeles, CA 90089, USA, e-mail: {\tt caire@usc.edu}.}
\thanks{The work by A. Guill\'en i F\`abregas work has been supported in part by the Australian Research Council under ARC grants DP0558861 and RN0459498.}}
\date{\today}
\begin{document}

\maketitle

%%%%%%%%%%%%%%%%%%%%%%%%%%%%%%%%%%%%%%%%%
%\vspace{-15mm}
\begin{abstract}
%\vspace{-3mm}
We study the problem of constructing coded modulation schemes over multidimensional signal sets in Nakagami-$m$ block-fading channels. In particular, we consider the optimal diversity reliability exponent of the error probability when the multidimensional constellation is obtained as the rotation of classical complex-plane signal constellations. We show that multidimensional rotations of full dimension achieve the optimal diversity reliability exponent, also achieved by Gaussian constellations. Multidimensional rotations of full dimension induce a large decoding complexity, and in some cases it might be beneficial to use multiple rotations of smaller dimension. We also study the diversity reliability exponent in this case, which yields the optimal rate-diversity-complexity tradeoff in block-fading channels with discrete inputs.
\end{abstract}

{\keywords Block-fading channels, diversity, linear rotations, maximum distance-separable (MDS) codes, outage probability.}

\newpage
%%%%%%%%%%%%%%%%%%%%%%%%%%%%%%%%%%%%%%%%%
%\vspace{-5mm}
\section{Introduction}
Rotated multidimensional constellations in fading channels were proposed in \cite{battail1989rrc,belfiore_ciss92} as a way of achieving high reliability with {\em uncoded} modulation in fading channels. Since, rotated constellations have been extensively studied, and have been shown to be an effective technique to achieve full-rate and full-diversity transmission in fading channels \cite{boutros1998ssd,BOV04,FNT04,frederique_thesis}. Traditionally, rotated constellations have always been studied uncoded, with the exception of some recent works for the multiple-input multiple-output (MIMO) channel \cite{nico_isit2004,kraidy8itv}.

In this work, we study the problem of constructing general coded modulation schemes over multidimensional signal sets, obtained by rotating classical complex-plane signal constellations, for block-fading channels with $B$ fading blocks (or degrees of freedom) per codeword \cite{ozarow_shamai_wyner}. The block-fading channel is a useful model for transmission over slowly varying fading channels, such as orthogonal frequency division multiplexing (OFDM) or slow time-frequency-hopped systems such as GSM or EDGE.

Despite the elegance of full-diversity rotations of dimension $B$, they induce large decoding complexity since the set of candidate points for detection at a given time instant is exponential with $B$. In fact, when uncoded rotations are used, the sphere decoder \cite{viterbo_boutros_sd} is usually employed to avoid exhaustive search over all candidate points. However, when coded modulation is used, the code itself can help to achieve full diversity. This means that sometimes rotations of smaller dimension $N<B$ might be sufficient. Also in the coded case, soft information should be provided to the decoder and this further complicates the problem. As a matter of fact, despite the recent advances in soft-output sphere decoding techniques \cite{boutros_globecom2003}, most of the proposed techniques still show performance limitations, which might be undesirable in practice. Therefore, in practice, one might be interested in using rotations of dimension smaller than $B$, in order to establish the tradeoff between diversity, rate, constellation size and {\em complexity} induced by the rotations. 

In this correspondence, we study the {\em reliability exponent}, namely, the optimal exponent of the error probability of such schemes with the signal-to-noise ratio (SNR) in a logarithmic scale, and illustrate the rate-diversity-complexity tradeoff for coded modulation schemes constructed over multidimensional signal sets.

%%%%%%%%%%%%%%%%%%%%%%%%%%%%%%%%%%%%%%%%%
\section{System Model}

We consider a single-input single-output block-fading channel with $B$ fading blocks, whose system model is given by the following,
\beq
\yv_b = \sqrt{\snr}\, h_b \,\xv_b + \zv_b \;\;\;\;\;\; b=1,\dotsc,B
\label{eq:model}
\eeq
where $h_b\in \CC$ is the $b$-th fading coefficient, $\yv_b\in\CC^{L}$ is the received signal vector corresponding to fading coefficient $b$, $\xv_b\in\CC^{L}$ is the portion of codeword allocated to block $b$ and $\zv_b\in\CC^{L}$ is the vector of i.i.d. noise samples $\sim \Nc_\CC(0,1)$. We assume that the transmitted signal is normalized in energy, i.e., $\EE[|x|^2]=1$. Hence, $\snr$ is the average received SNR.

We assume that the fading coefficients are i.i.d. from block to block and from codeword to codeword, and that they are perfectly known at the receiver, i.e, perfect channel state information at the receiver (CSIR). Since the channel coefficients are perfectly known to the receiver, we assume that the phase of the fading has been corrected. We also assume that the magnitudes of the channel coefficients follow a Nakagami-$m$ distribution% and thus
\[
p_{|h|}(\xi) = \frac{2m^m\xi^{2m-1}}{\Gamma(m)} e^{-m \xi^2}
\]
for $m>0$ \footnote{The literature usually considers $m\geq0.5$ \cite{proakis}. However, the distribution is well defined and reliable communication is possible for $0<m<0.5$.}
where
$\Gamma(\xi) \eqdef \int_0^{+\infty} t^{\xi-1} e^{-t} dt$
is the Gamma function \cite{abramowitz_stegun}. By analyzing Nakagami-$m$ fading, we are able to characterize a large class of fading statistics, including
Rayleigh fading by setting $m = 1$ and Rician fading with parameter $\Kc$ by setting $m =(\Kc + 1)^2/(2\Kc + 1)$ \cite{simon2000dco}. For future use we define $\gamma_b\eqdef|h_b|^2$, $b=1,\dotsc,B$.
We can express \eqref{eq:model} in matrix form as
\beq
\Ym = \sqrt{\snr}\, \Hm \,\Xm + \Zm
\label{eq:model_matrix}
\eeq
where $\Ym = [\yv_1 ,\dotsc, \yv_B]^T \in \CC^{B\times L},~\Xm = [\xv_1 ,\dotsc, \xv_B]^T=[\Xm_1,\dotsc,\Xm_{L}] \in \CC^{B\times L}, ~\Zm = [\zv_1 ,\dotsc, \zv_B]^T \in \CC^{B\times L}$ and $\Hm = \diag(h_1,\dotsc,h_B)\in \CC^{B\times B}$.

\begin{figure}[htbp]
  \centering 
  \includegraphics[width=1\columnwidth]{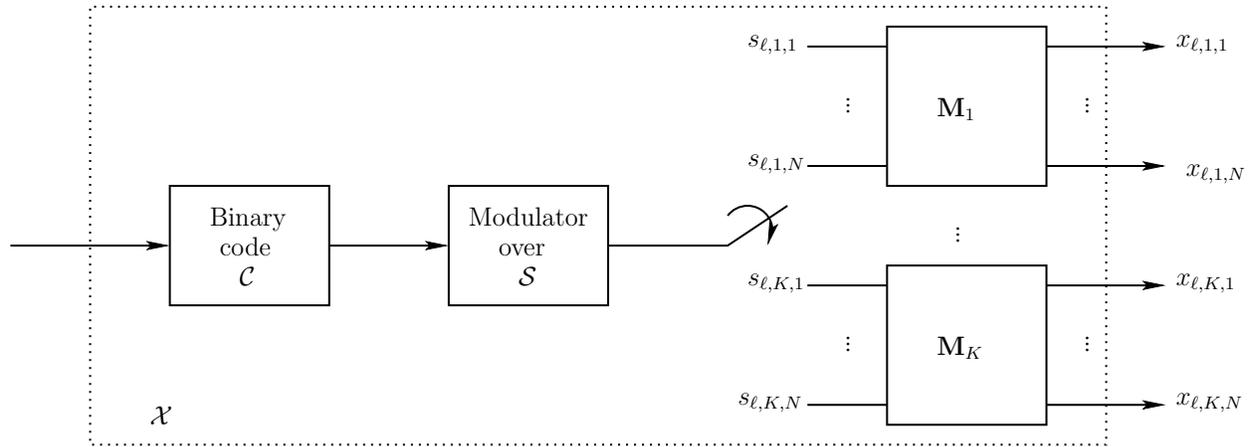}
  \caption{Block diagram for coded modulation with $K$ rotated constellations with rotation matrices $\Mm_1,\dotsc,\Mm_K$.}
  \label{fig:code_modulation_rotation}
\end{figure}

We consider that codewords $\Xm$ form a coded modulation scheme $\Xc\subset \CC^{B\times L}$. In particular, we consider that $\Xc$ is obtained as the concatenation of a binary code $\Cc\in\FF^n_2$ of rate $r$, a modulation over the signal constellation $\Sc\in\CC$ with $M=\log_2|\Sc|$, and $K$ rotations $\Mm_k\in\CC^{N\times N}$ with $KN= B$ (see Figure \ref{fig:code_modulation_rotation}). In particular we have that at channel use $\ell=1,\dotsc,L$
\vspace{-3mm}
\beq
\xv_{\ell,k} = \Mm_k \sv_{\ell,k}
\vspace{-3mm}
\eeq
where $\sv_{\ell,k}=(s_{\ell,k,1},\dotsc,s_{\ell,k,N})^T\in\Sc^{N}$ is the vector of complex-plane signal constellation symbols that is rotated by the $k$-th rotation matrix, $\xv_{\ell,k}=(x_{\ell,k,1},\dotsc,x_{\ell,k,N})^T$ is the portion of transmitted signal at the $\ell$-th channel use that has been rotated by the $k$-th rotation, and
\[
\xv_{\ell} = [\xv_{\ell,1}^T,\dotsc,\xv_{\ell,K}^T]^T 
\]
is the transmitted signal at the $\ell$-th channel use. The rotation matrices are constrained to be unitary, i.e., $\Mm_k\Mm_k^\dag=\Id$. We will be interested in {\em full-diversity} rotations, namely, rotation matrices $\Mm$ for which $\forall \sv,\sv'\in\Sc^{N}, \sv\neq\sv'$
\vspace{-3mm}
\beq
\Mm(\sv-\sv') \neq \zerov
\vspace{-3mm}
\eeq
componentwise. This implies that, if the vector $\sv-\sv'$ has any number of non-zero components, its rotated version $\Mm(\sv-\sv')$ will have {\em all} non-zero components. In this paper we will use some specific full-diversity matrices of dimension $N=2$ and $N=4$. For the sake of completeness, we report the corresponding matrices in the following. The reader is referred to \cite{BOV04,FNT04,frederique_thesis,rotations_web} for information on how these matrices have been designed. The $N=2$ cyclotomic rotation matrix is given by \cite{rotations_web}
\beq
\Mm = \left(\begin{matrix}
-0.5257311121 &-0.8506508083\\
-0.8506508083  & 0.5257311121
\end{matrix}\right).\notag
\eeq
The $N=4$ Kr\"uskemper rotation matrix is given by \cite{rotations_web}
\beq
\Mm = \left(\begin{matrix}
 -0.3663925121 &  -0.2264430248 & -0.474464708 & -0.7677000246\\
  -0.7677000238 & -0.4744647078  & 0.2264430248   &0.3663925106\\
   0.4230815704  &-0.6845603618  &-0.5049593144  & 0.3120820189\\
   0.3120820187  &-0.5049593142   &0.6845603618 & -0.4230815707
   \end{matrix}\right).\notag
\eeq
The $N=4$ mixed rotation matrix is given by \cite{rotations_web}
\beq
\Mm = \left(\begin{matrix}
0.2011885864868&   0.3255299710843 &  0.284523627604  & 0.4603689000663\\
   0.3255299710843 & -0.2011885864868  & 0.4603689000663 & -0.284523627604\\
   0.4857122140913 &  0.7858988711506  &-0.6869008005781 & -1.1114288422349\\
   0.7858988711506 & -0.4857122140913 & -1.1114288422349  & 0.6869008005782\\
  \end{matrix}\right).\notag
\eeq
Reference \cite{rotations_web} reports rotation matrices using the row convention used in \cite{conway_sloane}. In this paper, we use a column convention for lattice generator matrices, and therefore, matrices from \cite{rotations_web} are transposed.

The rate in bits per channel use of this scheme is independent of $N$, and is given by $R=rM$. This general formulation includes the case where only one single rotation of dimension $B$ is used, as well as the other extreme, with $B$ trivial rotations of dimension $N=1$ (the non-rotated case). As we shall seen in the following, although the rate is independent of $N$, the reliability exponent does depend on $N$. 

\begin{definition}
The block-diversity of a coded modulation scheme $\Xc \subset \CC^{B\times L}$ is defined as
\beq
\delta = \min_{\substack{\Xm(i),\Xm(j) \in \Xc\\ j \neq i}} |\{b\in(1,\dotsc,B) \;\;|\;\; \xv_b(i) \neq \xv_b(j)\}|.
\eeq
\label{def:block_diversity}
\end{definition}
In words, the block diversity is the minimum number of nonzero rows of $\Xm(i)-\Xm(j)$ for any pair of codewords $\Xm(j) \neq \Xm(i)\in \Xc$.
\begin{proposition}
Given a coded modulation scheme $\Xc \subset \CC^{B\times L}$, the block diversity is upperbounded by
\beq
\delta \leq N\left(1+\left\lfloor \frac{B}{N}\left(1-\frac{R}{M}\right) \right\rfloor\right).
\label{eq:sb_block_div}
\eeq
\label{prop:sb_ub}
\end{proposition}
\noindent\begin{proof}
The result follows from the straightforward application of the Singleton bound to the coded modulation $\Xc$ seen as a code of block-length $K$, over an alphabet of size $2^{M N L}$.
\end{proof}

We will say that a code is blockwise maximum-distance separable (MDS) if it attains the Singleton bound of Proposition \ref{prop:sb_ub} with equality.

%%%%%%%%%%%%%%%%%%%%%%%%%%%%%%%%%%%%%%%%%%%%%%%%%%%%%%%%%%%
\section{Outage Probability}
Strictly speaking, the channel defined in \eqref{eq:model} is not information stable and has zero capacity for any finite $B$ \cite{verdu_han}, since there is a non-zero probability that the transmitted message is detected in error even for codes of infinite length. For sufficiently large $L$, the word error probability $P_e(\snr,\Xc)$ of any coding scheme $\Xc \subset \CC^{B\times L}$ is lowerbounded by the {\em information outage probability} \cite{ozarow_shamai_wyner,biglieri_proakis_shamai}, given by
\beq
P_e(\snr,\Xc) \geq \Pout \eqdef \Pr(I(\snr,\Hm) \leq R).
\eeq
where $I(\snr,\Hm)$ is the input-output mutual information of the channel for a given fading realization $\Hm$. In this work, we will study the behavior of $\Pout$ for large $\snr$, for which the optimal power allocation when no CSI is available at the transmitter, corresponds to evenly distributing the available power across all $B$ blocks. In the case of uniform allocation, and for a fixed $\Hm$, the outage probability is minimized when the entries of $\Xm\in\Xc$ are i.i.d. Gaussian $\sim\Nc_{\CC}(0,1)$. In this case \cite{cover_thomas}
\beq
I(\snr,\Hm) = \frac{1}{B}\sum_{b=1}^B \log_2(1+\snr\gamma_b).
\eeq
When the coded modulation scheme shown in Figure \ref{fig:code_modulation_rotation} is used (assuming uniform inputs), we can express the {\em instantaneous mutual information} in bits per channel use for a given channel realization $\Hm$ as
\beq
I(\snr,\Hm) = \frac{1}{K}\sum_{k=1}^K \frac{1}{N} I_k(\snr,\Hm_k) = \frac{1}{B}\sum_{k=1}^K I_k(\snr,\Hm_k) \nonumber
\eeq
where the mutual information of the $N\times N$ MIMO channel induced by the $k$-th rotation is (see e.g., \cite{hochwald_tenbrink,Caire/Colavolpe:2001} for the derivation of the mutual information of discrete-input MIMO channels)
\begin{align}
I_k(\snr,\Hm_k) = MN - \frac{1}{2^{MN}}\sum_{\sv\in\Sc^{N}} \EE_\zv\left[\log_2 \left(1+\sum_{\sv'\neq\sv} e^{- \|\sqrt\snr\,\Hm_k\Mm_k(\sv-\sv') + \zv\|^2 + \|\zv\|^2}\right) \right]
\label{eq:mi_rotation}
\end{align}
and $\Hm_k=\diag(h_{(k-1)N+1},\dotsc,h_{kN})\in\CC^{N\times N}$ are the channel coefficients {\em used} by rotation $k$, and $\zv\in\CC^N$ is a dummy AWGN vector over which the expectation is computed. For small $N$, the expectation over the noise vector $\zv$ in \eqref{eq:mi_rotation} can be efficiently computed using the Gauss-Hermite quadrature rules \cite{abramowitz_stegun}.

Note that concatenating a Gaussian random code with a rotation of dimension $B$ brings no benefit in terms of exponent nor mutual information. In fact, the output of the rotated Gaussian i.i.d. vector is also a Gaussian i.i.d. vector with identical distribution, provided that the rotation matrix is unitary. Therefore, the mutual information
\begin{align}
I(\snr,\Hm) &= \frac{1}{B}\log_2 \det \left(\Id + \snr \, \Hm\Mm \Mm^\dag \Hm^\dag \right )= \frac{1}{B} \sum_{b=1}^B \log_2(1+\snr \gamma_b).
\end{align}
is the same than without rotation, and so is therefore the corresponding diversity exponent. Rotations are usually seen as information lossless, when in fact they are simply not needed when combined with Gaussian inputs.

\begin{figure}[htbp]
  \centering 
	\includegraphics[width=1\columnwidth]{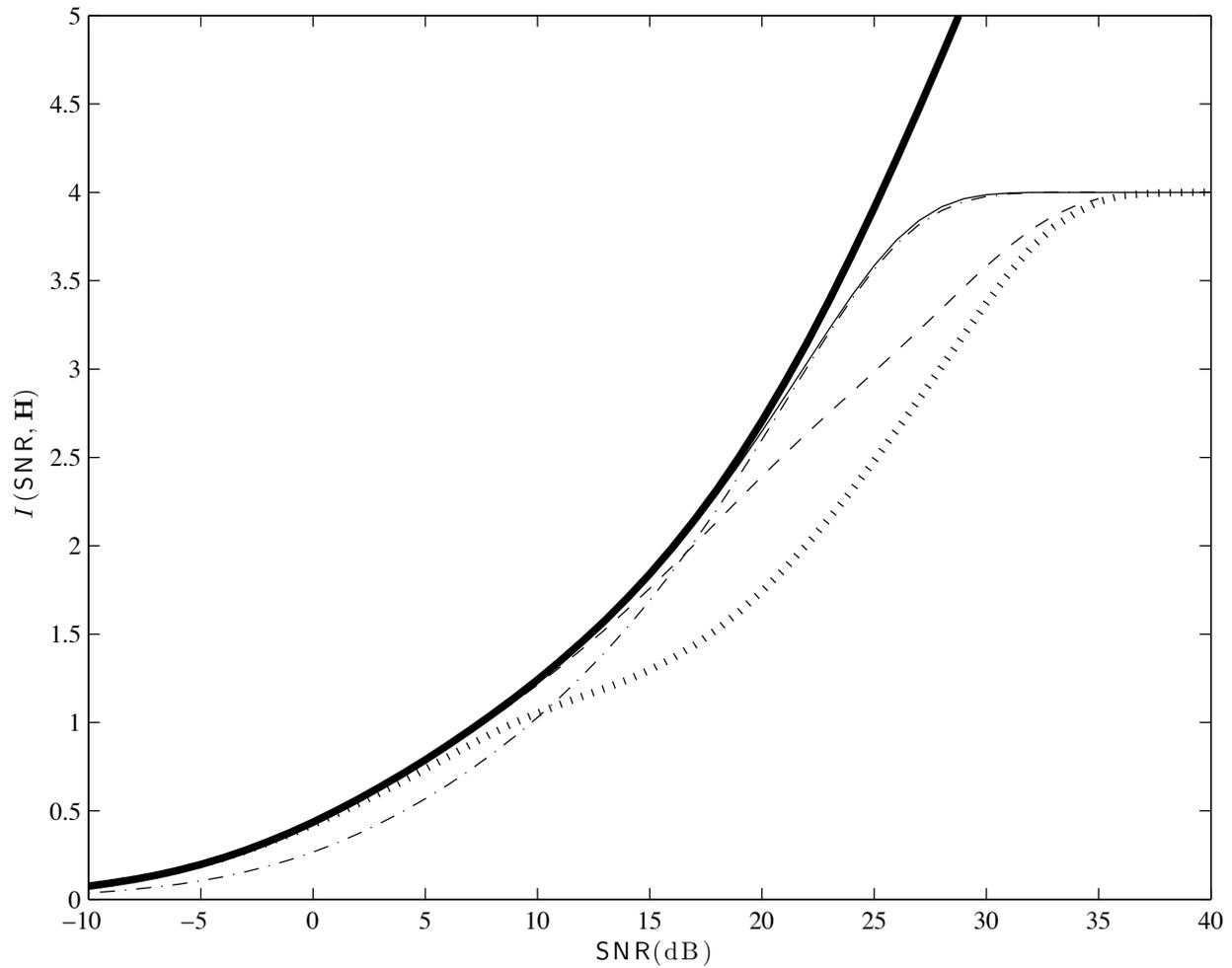}
  \caption{Instantaneous mutual information $I(\snr,\Hm)$ (bits/channel use) in a block-fading channel with $B=4$ blocks and $h_1 = 1.5$ and $h_2=h_3=h_4=0.1$ with Gaussian inputs (thick solid) and rotated $16$-QAM inputs with the optimal Kr\"uskemper (thin solid), mixed (thin dash-dotted), $2$ independent $2$-dimensional cyclotomic rotations (thin dashed) and no rotations (thick dotted). }
  \label{fig:mi_gauss_qam_rotations}
\end{figure}

Figure \ref{fig:mi_gauss_qam_rotations} shows the mutual information with Gaussian inputs, unrotated $16$-QAM (identity rotation) and rotated $16$-QAM in a block-fading channel with $B=4$ blocks and $h_1 = 1.5$ and $h_2=h_3=h_4=0.1$. This choice of the channel coefficients is particularly interesting since $3$ out of the $4$ components are in a deep fade \footnote{Note that in this nonergodic scenario, the ergodic information rate averaged over the channel realizations does not have a practical relevance. Instead, we are interested in finding out the behavior of the system for {\em bad} channels which dominate the outage probability for large $\snr$.}. Rotations of dimension $N$ yield vanishing (for large $\snr$) error probability whenever there are up to $N-1$ deeply faded blocks  \cite{boutros1998ssd,BOV04,FNT04,frederique_thesis}. The mutual information achieved by the rotated $16$-QAM is very close to that attained by the Gaussian distribution for a range of $\snr$ significantly wider than unrotated $16$-QAM. For example, at $\snr = 25$dB, the Kr\"uskemper rotation gains $1$ bit of information with respect to unrotated $16$-QAM. Combining $2$ cyclotomic rotations of dimension $N=2$ brings also significant information gains with respect to unrotated $16$-QAM. As we shall see, this effect brings substantial exponent benefits with respect to the unrotated case. We also appreciate some difference between optimal Kr\"uskemper and the mixed ($2\times 2$) rotations, especially at low rates. As a matter of fact, rotations provide only mutual information advantages at high rates. At low rates, unrotated transmission performs almost as well with much less decoding complexity. 

%%%%%%%%%%%%%%%%%%%%%%%%%%%%%%%%%%%%%%%%%%%%%%%%%%%%%%%%%%%%%%%%
\section{Optimal Reliability}

We define the {\em diversity reliability exponent} of a given coded modulation scheme $\Xc$ as
\beq
d_\Xc = \lim_{\snr \to +\infty}-\frac{\log P_e(\snr,\Xc)}{\log\snr}
\eeq 
and the optimal diversity reliability exponent is
\beq
d^\star \eqdef \sup_\Xc d_\Xc=\sup_\Xc\lim_{\snr \to +\infty}-\frac{\log P_e(\snr,\Xc)}{\log\snr}.
\eeq

When no particular structure is imposed on the coded modulation scheme $\Xc$, we have the following result.
\begin{lemma}
\label{lemma:gi}
The diversity reliability exponent $d_\Xc$ of any coded modulation scheme $\Xc$ subject to the power constraint $\frac{1}{BL}\EE[\|\Xm\|^2]\leq 1$ is upperbounded by
\beq
d_\Xc \leq d^\star = mB.
\label{eq:exponent_opt}
\eeq
The optimal diversity reliability exponent can be achieved by random Gaussian codes of rate $R>0$ with entries $\sim\Nc_\CC(0,1)$. The optimal exponent $d^\star$ can also be achieved by random coded modulation schemes $\Xc$ of rate $R$ consisting of a random coded modulation scheme over a discrete signal constellation $\Sc$ of size $|\Sc|=2^M$ concatenated with a full-diversity rotation of dimension $B$, whenever $0\leq\frac{R}{M}<1$.
\end{lemma}
\begin{proof}
The converse is proved in \cite{albert_beppe_it, nguyen2007}. Furthermore, \cite{albert_beppe_it, nguyen2007} also show that the random Gaussian ensemble achieves the optimal exponent. What is left to prove is that the random coded modulation scheme over a single full-diversity rotation of dimension $B$ achieves the same exponent. This is proved in Appendix \ref{appendix:proof_prop_lb}, by letting $N=B$.
\end{proof}

We have included the achievability with the random coded modulation ensemble over the $B$-dimensional rotated constellation to illustrate that a coding scheme with discrete inputs can also achieve the optimal exponent. This result which is based on a {\em divide and conquer} approach, should be rather intuitive: the rotation of dimension $B$ takes care of achieving full diversity while the coding gain is then left to the outer coded modulation scheme over $\Sc$. 
When no rotations are used, the optimal diversity reliability exponent is given by the Singleton bound \cite{nguyen2007}
\beq
d^\star = m\left(1 + \left\lfloor B\left(1 - \frac{R}{M}\right)\right\rfloor\right).
\label{eq:sb}
\eeq

\begin{figure}[htbp]
  \centering 
  \includegraphics[width=1\columnwidth]{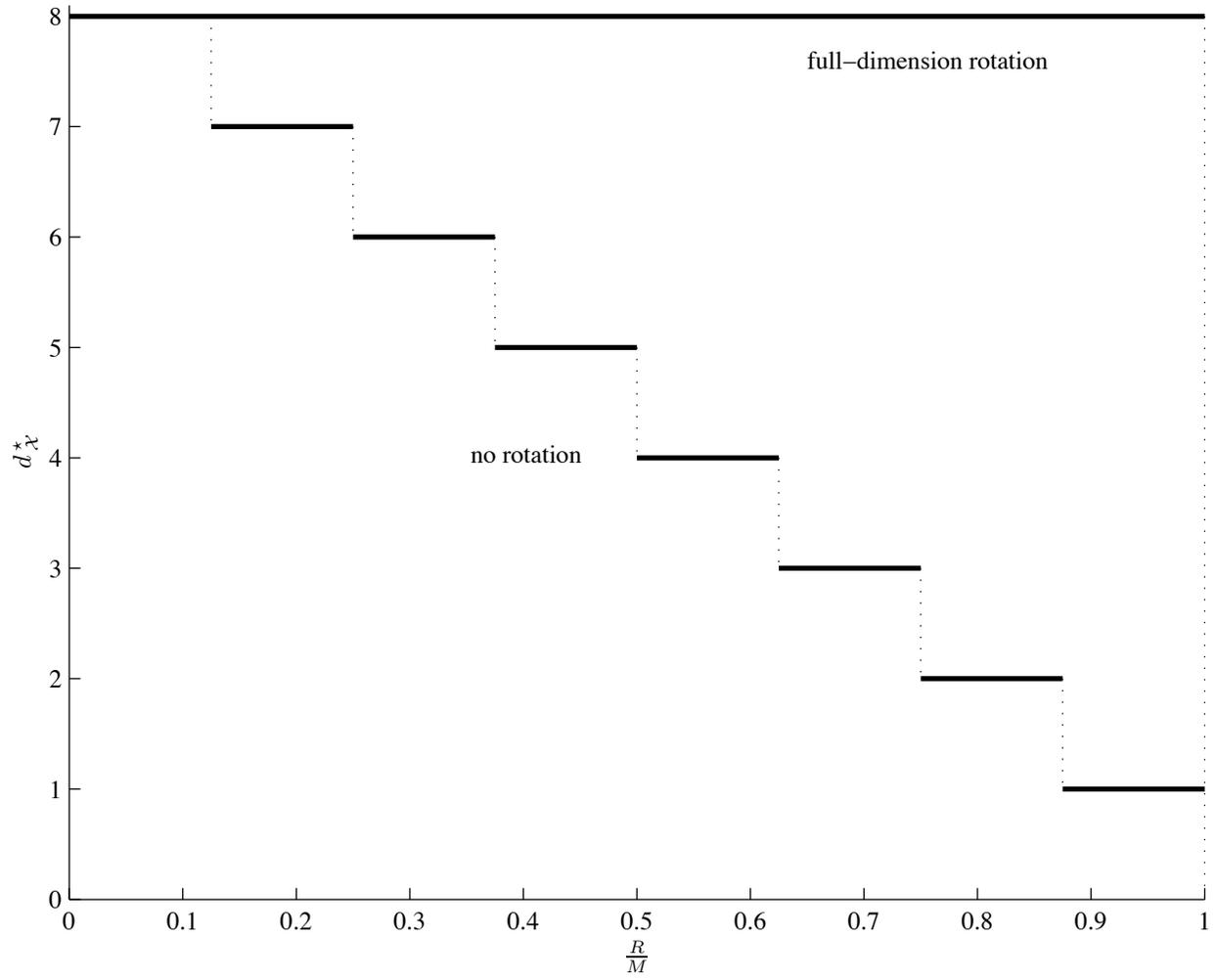}
  \caption{Diversity reliability exponents for $B=8$ and $m=1$. Optimal exponent \eqref{eq:exponent_opt} and Singleton bound \eqref{eq:sb}.}
  \label{fig:exponents_b8_opt_di}
\end{figure}

As shown in Figure \ref{fig:exponents_b8_opt_di} the advantage of rotations is clear: they can achieve the optimal diversity reliability exponent for the whole range of rates. Instead, when no rotations are used, the largest rate such that optimal diversity reliability exponent is achieved is $R=\frac{M}{B}$.

As outlined in the Introduction, full-diversity rotations induce large decoding complexity, since the size of the set of candidate points at a given time instant is $2^{MB}$. We are therefore interested in characterizing the optimal diversity reliability exponent when rotations of smaller size $N<B$ are employed. We have the following results
\begin{proposition}
\label{prop:exponent_singleton_lattice_ub}
The diversity reliability exponent for the coded modulation schemes based on $K$ rotations of dimension $N$, in a Nakagami-$m$ block-fading channel with $B=KN$ blocks is upperbounded by
\beq
d_\Xc \leq  mN\left(1+\left\lfloor \frac{B}{N}\left(1-\frac{R}{M}\right) \right\rfloor\right).
\eeq
\end{proposition}
\begin{proof}
See Appendix \ref{appendix:proof_prop_ub}.
\end{proof}

\begin{proposition}
\label{prop:exponent_singleton_lattice_lb}
The diversity reliability exponent in a Nakagami-$m$ block-fading channel with $B=KN$ of random coded modulation schemes based on $K$ rotations of dimension $N$ of length $L$ satisfying $\lim_{\snr\to\infty}\frac{L}{\snr}=\lambda$, is lowerbounded by 
\beq
d_\Xc \geq \begin{cases}
\lambda BM\log2 \left(1-\frac{R}{M}\right) & \text{if $0\leq\lambda NM\log2<m$}\\\\
\min\Biggl\{m N\left\lceil \frac{B}{N}\left(1-\frac{R}{M}\right) \right\rceil  ~,~ m N \left\lfloor \frac{B}{N}\left(1-\frac{R}{M}\right)\right\rfloor\\
+ \lambda M\log2\left( B\left(1-\frac{R}{M}\right) - N \left\lfloor \frac{B}{N}\left(1-\frac{R}{M}\right)\right\rfloor\right)\Biggr\} & \text{otherwise}.
\end{cases}
\eeq
%
%\beq
%d_\Xc \geq  \lambda BM\log2 \left(1-\frac{R}{M}\right)
%\eeq
%when $0\leq\lambda NM\log2<m$ and by
%\begin{align}
%&d_\Xc \geq \min\Biggl\{m N\left\lceil \frac{B}{N}\left(1-\frac{R}{M}\right) \right\rceil  , m N \left\lfloor \frac{B}{N}\left(1-\frac{R}{M}\right)\right\rfloor \nonumber\\
%&+ \lambda M\log2\left( B\left(1-\frac{R}{M}\right) - N \left\lfloor \frac{B}{N}\left(1-\frac{R}{M}\right)\right\rfloor\right)\Biggr\}
%\end{align}
%otherwise.
\end{proposition}
\begin{proof}
See Appendix \ref{appendix:proof_prop_lb}.
\end{proof}

The proof of the last two Propositions closely follows the reasoning of \cite{albert_beppe_it, nguyen2007}. Although the basic steps of the proofs are the same, the inclusion of the rotation matrix of dimension $N$ is nontrivial, and a detailed proof is needed to track the impact of the rotation dimension $N$ in the final expression of the resulting exponent.

The preceding results lead to the following Theorem.
\begin{theorem}
\label{th:exponent_singleton_lattice}
The optimal diversity reliability exponent for the coded modulation schemes based on $K$ rotations of dimension $N$, in a Nakagami-$m$ block-fading channel with $B=KN$ blocks is given by
\beq
d_\Xc^\star = mN\left(1+\left\lfloor \frac{B}{N}\left(1-\frac{R}{M}\right) \right\rfloor\right)
\eeq
whenever $\frac{B}{N}\left(1-\frac{R}{M}\right)$ is not an integer.
\end{theorem}
\begin{proof}
Proposition \ref{prop:exponent_singleton_lattice_ub} shows that
\beq
d_\Xc \leq  mN\left(1+\left\lfloor \frac{B}{N}\left(1-\frac{R}{M}\right) \right\rfloor\right).
\eeq
Letting $\lambda\to\infty$ in Proposition \ref{prop:exponent_singleton_lattice_lb} shows that
\beq
d_\Xc \geq  mN\left\lceil \frac{B}{N}\left(1-\frac{R}{M}\right) \right\rceil.
\eeq
Noting that $\lceil x \rceil = \lfloor x \rfloor +1$ whenever $x$ is not an integer leads the desired result.
\end{proof}

As we observe, Theorem \ref{th:exponent_singleton_lattice} gives a dual result to that of \cite{nguyen2007} and shows that the optimal exponent is given by $m$ times the Singleton bound of \eqref{eq:sb_block_div}, proving its optimality and separating the roles of the channel distribution (through $m$) and of the code construction. 
%When $N=1$, there is no rotation, and the inputs to the channel are directly signal constellation points drawn from $\Sc$. In this case, we have that the optimal reliability exponent is given by the Singleton bound \cite{KnoppHumblet,albert_beppe_it}
%\beq
%d_\Xc^\star = m \left(1+\left\lfloor B\left(1-\frac{R}{M}\right) \right\rfloor\right)
%\eeq
%for any $R\leq M$, and 
The optimal codes are {\em blockwise} MDS in a channel with $B$ blocks. 
For $N>1$, Theorem \ref{th:exponent_singleton_lattice} suggests that the optimal coding scheme is to use a coded modulation scheme constructed over $\Sc$  which is MDS in a block-fading channel with $K=\frac{B}{N}$ blocks concatenated with rotations of dimension $N$. In this case the MDS constraint on the code is relaxed, since it has to be MDS for a smaller number of blocks, at an expense of a decoding complexity increase. Theorem \ref{th:exponent_singleton_lattice} implicitly introduces an equivalent channel model, namely, a block-fading channel with $K=\frac{B}{N}$, where each block has diversity $mN$. When $K=1$, $N=B$, there is only one single rotation of {\em full} dimension, Theorem \ref{th:exponent_singleton_lattice} generalizes Lemma \ref{lemma:gi}. The optimal coding scheme here does not need to be MDS. Therefore, Theorem \ref{th:exponent_singleton_lattice} generalizes and proves the optimality of the {\em modified Singleton bound} introduced in \cite{nico_isit2004}. 

\begin{figure}[htbp]
  \centering 
  \includegraphics[width=1\columnwidth]{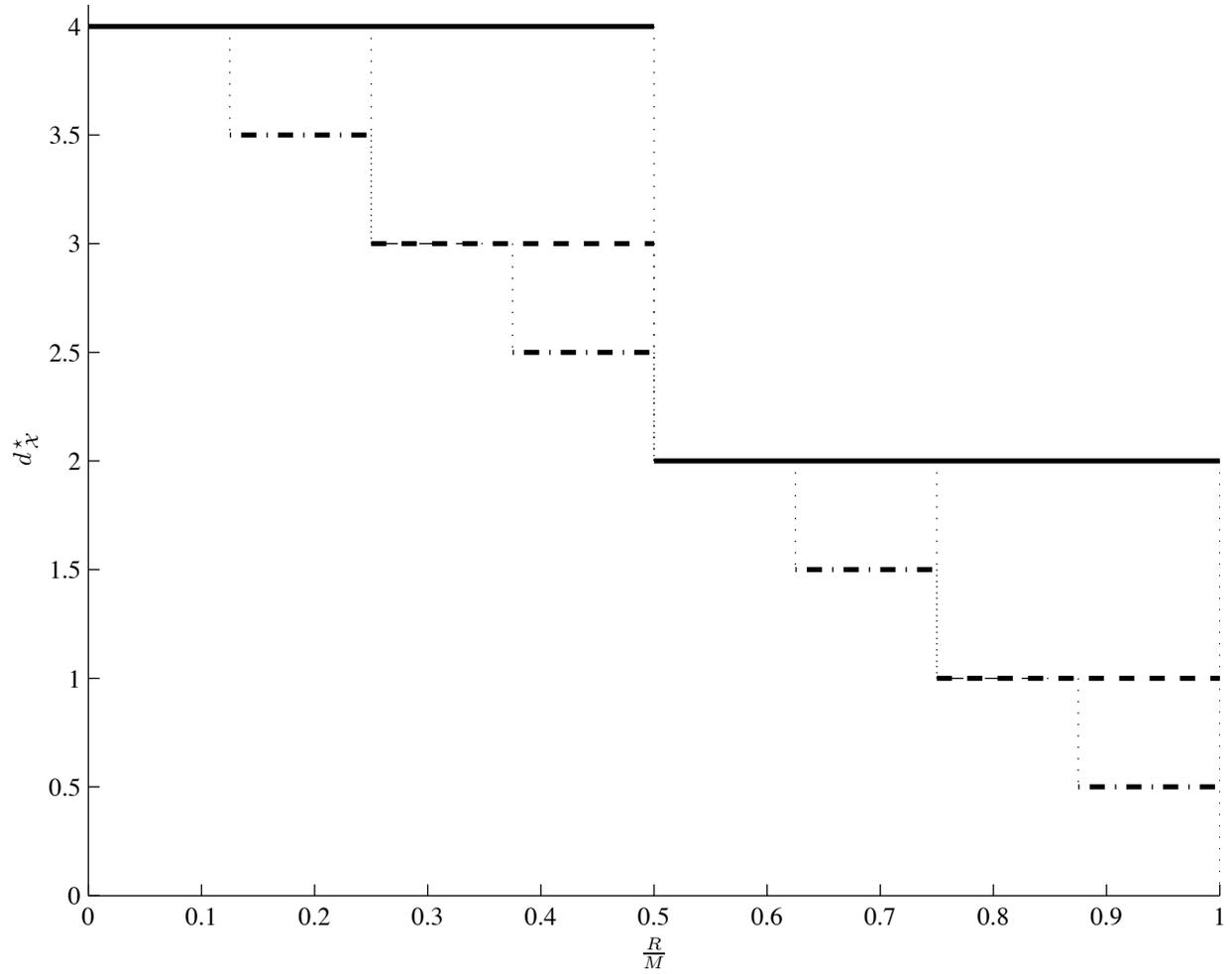}
  \caption{Reliability exponents for $B=8$, $m=0.5$ and rotations of dimensions $N=1$ (dash-dotted), $N=2$ (dashed) and $N=4$ (solid).}
  \label{fig:exponents_b8}
\end{figure}

\begin{figure}[htbp]
  \centering 
  \includegraphics[width=1\columnwidth]{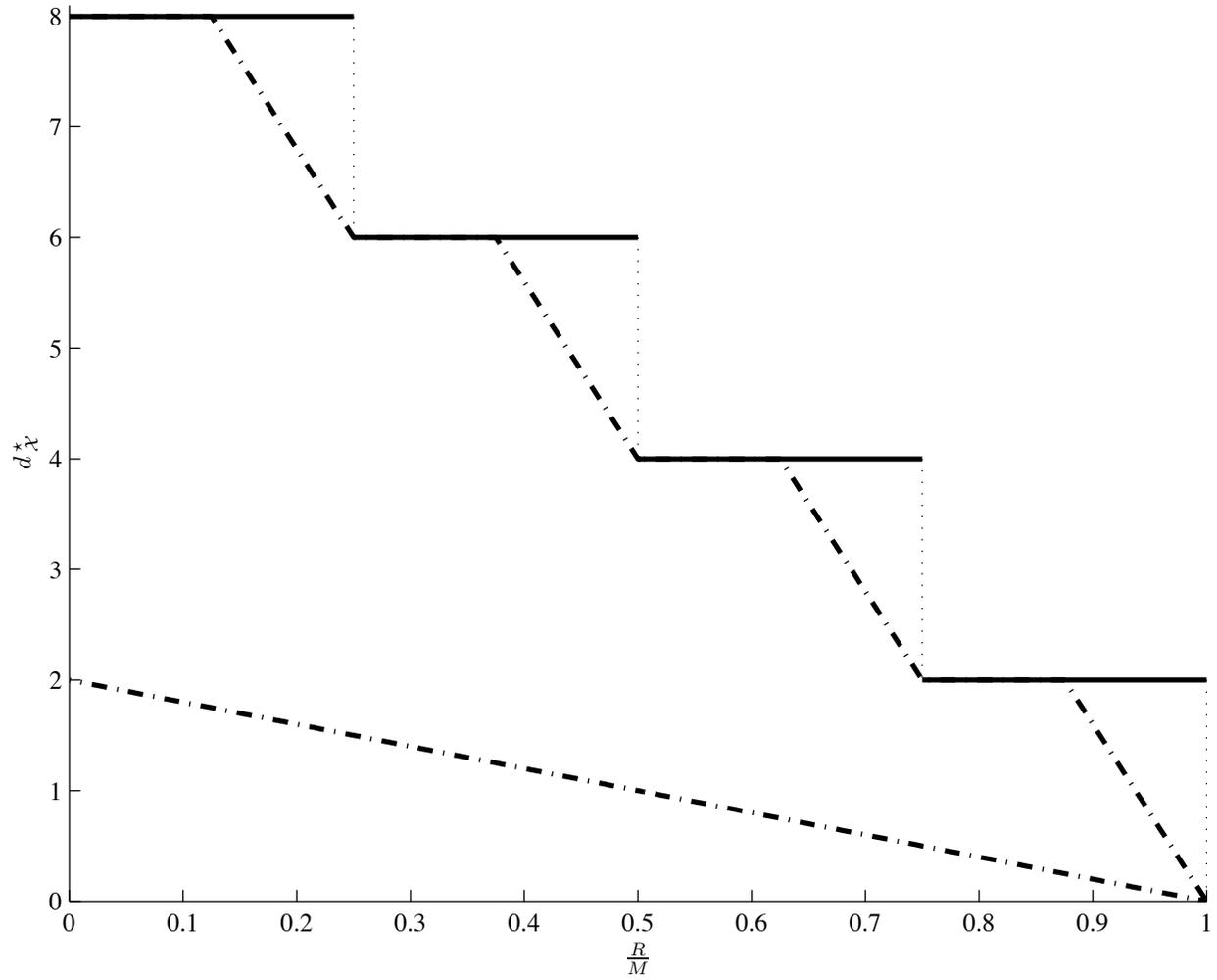}
  \caption{Reliability exponents for $B=8$, $m=1$ and rotations of dimensions $N=2$. The random coding exponents for $\lambda M \log2=\frac{m}{2N}$ (lower dash-dotted curve) and $\lambda M \log2=\frac{4m}{N}$ (upper dash-dotted curve) are also shown.}
  \label{fig:exponents_b8_n2}
\end{figure}

Figure \ref{fig:exponents_b8} shows the reliability exponents in the case of $B=8$, $m=0.5$ and $N=1,2,4$. The figure confirms the intuition behind such designs that the rotations should increase the reliability exponent. For example, for $\frac{R}{M}=\frac{1}{2}$, we have that with classical complex-plane inputs the reliability exponent is $d_\Xc^\star=m5$, while for rotations with $N=2$ the exponent is $d_\Xc^\star=m6$ and for $N=4$ the exponent is $d_\Xc^\star=m8$, full diversity. This approach can be seen as a divide-and-conquer approach, namely, the task of achieving diversity is split between both, the code $\Cc$ and the rotations. Figure \ref{fig:exponents_b8_n2} shows the diversity upper bound as well as the random coding lower bounds given in Propositions \ref{prop:exponent_singleton_lattice_ub} and \ref{prop:exponent_singleton_lattice_lb}, respectively. As we see, if $\lambda$ is increased, both bounds coincide in a larger support. Eventually, for $\lambda\to\infty$ they coincide wherever they are continuous.

\begin{figure}[htbp]
  \centering 
\includegraphics[width=1\columnwidth]{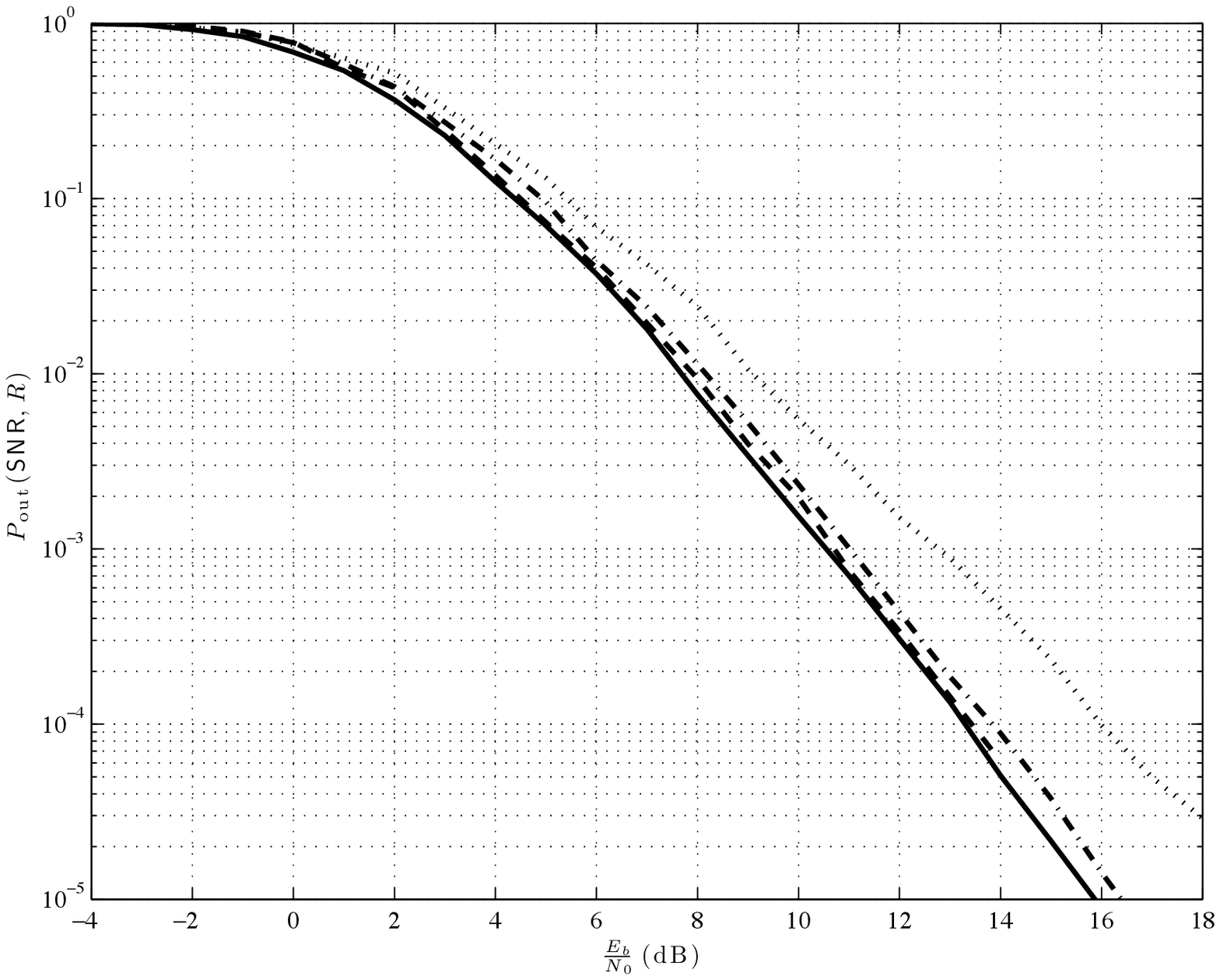}
 \caption{Outage probability for $R=1$ bits per channel use in a block-fading channel with $B=4$, $m=1$, with Gaussian (solid line), rotated QPSK inputs with one Kr\"uskemper rotation of dimension $N=4$ (dashed line), rotated QPSK inputs with two cyclotomic rotations of dimension $N=2$ (dash-dotted) and unrotated QPSK inputs (dotted).}
  \label{fig:pout_rotations_qpsk}
\end{figure}

\begin{figure}[htbp]
  \centering
  \includegraphics[width=1\columnwidth]{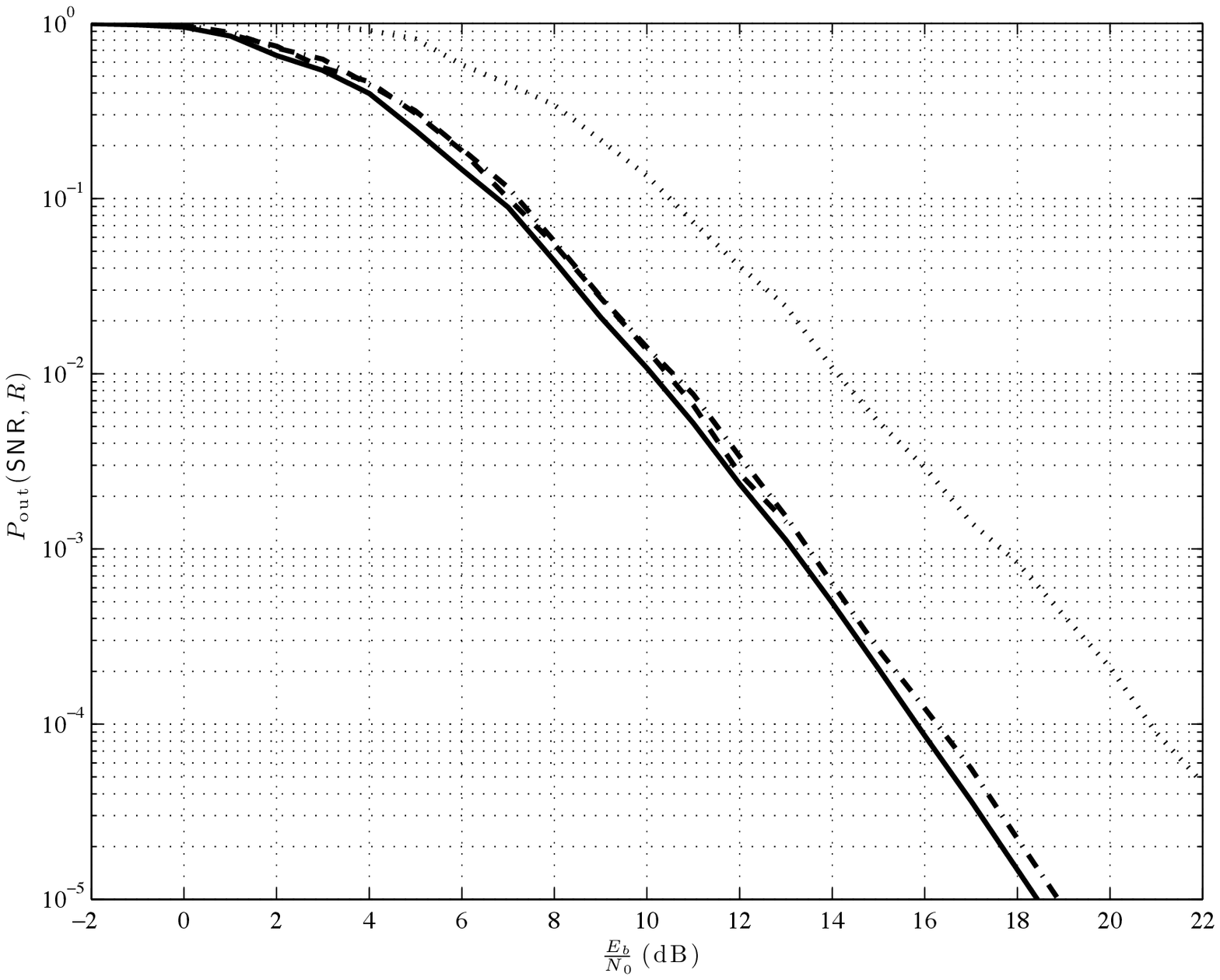}
 \caption{Outage probability for $R=2$ bits per channel use in a block-fading channel with $B=4$, $m=1$, with Gaussian(solid line), rotated $16$-QAM inputs with one Kr\"uskemper rotation of dimension $N=4$ (dashed line), rotated $16$-QAM inputs with two cyclotomic rotations of dimension $N=2$ (dash-dotted) and unrotated $16$-QAM inputs (dotted).}
  \label{fig:pout_rotations_qam}
\end{figure}

To illustrate the performance benefits of rotations, Figures \ref{fig:pout_rotations_qpsk} and \ref{fig:pout_rotations_qam} show $\Pout$ as a function of $\frac{E_{\rm b}}{N_0}$ in a block-fading channel with $m=1$ and $B=4$ for $R=2$, with Gaussian inputs (solid), discrete inputs (dotted), rotated discrete inputs with two cyclotomic rotations with $N=2$ (dash-dotted) and rotated discrete inputs with one Kr\"uskemper rotation with $N=4$ (dashed). Gaussian inputs achieve the optimal exponent, namely $d^\star=B=4$, while unrotated inputs have $d^\star_\Xc=3$ \cite{albert_beppe_it}. As we observe from the curves, using two rotations of dimension $N=2$, not only allows to recover the largest possible exponent (in agreement with Theorem \ref{th:exponent_singleton_lattice}) but also brings a large gain. Using a rotation of dimension $N=4$ incurs much larger complexity and does not bring any exponent or gain improvements.

\begin{figure}[htbp]
  \centering
  \includegraphics[width=1\columnwidth]{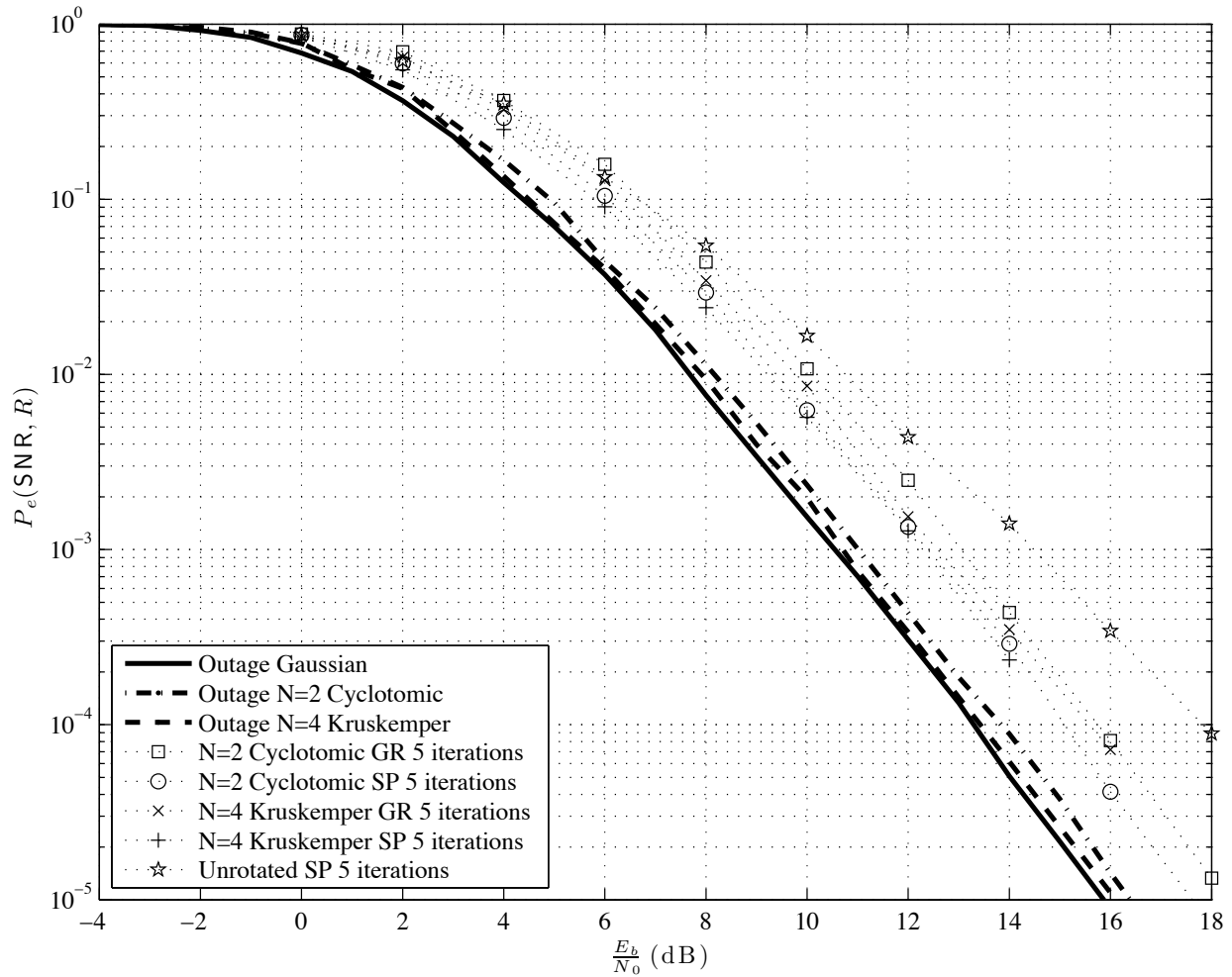}
 \caption{Error probability for $R=1$ bits per channel use in a block-fading channel with $B=4$, $m=1$ using the  $(5,7)_8$ convolutional code and QPSK modulation with Gray (GR) and set-partitioning (SP) labeling. The outage probabilities with Gaussian inputs (thick solid line), rotated QPSK inputs with one Kr\"uskemper rotation of dimension $N=4$ (dashed line), rotated QPSK inputs with two cyclotomic rotations of dimension $N=2$ (dash-dotted) are shown for reference. }
  \label{fig:sim_B4_QPSK}
\end{figure}

To illustrate that the above theoretical results are approachable with practical coding schemes, Figure \ref{fig:sim_B4_QPSK} shows the error probability of rotated and unrotated systems with QPSK modulation using the $(5,7)_8$ convolutional code with 128 information bits per frame. The outage probabilities with Gaussian inputs (thick solid line), rotated QPSK inputs with one Kr\"uskemper rotation of dimension $N=4$ (dashed line), rotated QPSK inputs with two cyclotomic rotations of dimension $N=2$ (dash-dotted) are shown for reference, as well as the performance of the unrotated scheme, whose corresponding outage probability has been removed for the sake of clarity. In the case of two rotations of dimension $N=2$, we separately use bit-interleaved coded modulation (BICM) \cite{CaireBICM} followed by a rotation on the outputs generated by generator polynoimial $5_8$ and $7_8$. Since the $(5,7)_8$ convolutional code has full-diversity in a block-fading channel with $K=2$ blocks, this blockwise operation allows the overall coding scheme to achieve full-diversity. A similar construction can be obtained using blockwise concatenated codes \cite{albert_beppe_it} or multiplexed turbo-codes \cite{boutros:tcd}. These coded modulation schemes will closely approach the outage probability of the channel for any (sufficiently large) block length. Rotated systems use exhaustive iterative decoders, i.e., we compute the metrics or all the candidate points \cite{hochwald_tenbrink}. Again, as we observe, the gain obtained by using rotations is significant. As a matter of fact, all systems using rotations show a steeper slope to that of the unrotated case. Furthermore, we observe that using a rotation of full dimension $N=4$ yields once more a small gain with respect to using two rotations of dimension $N=2$, while significantly increasing the decoding complexity. We also observe that, set-partitioning labeling yields some performance advantage over Gray labeling. From results not shown here, both Gray and set-partitioning show improved performance with the iterations. This is due to the the fact that rotations induce an equivalent MIMO channel, and the iterative decoder assists in iteratively removing the self-interference introduced by the rotation.

%%%%%%%%%%%%%%%%%%%%%%%%%%%%%%%%%%%%%%%%%
\section{Conclusions}
We have studied coded modulation schemes over Nakagami-$m$ block-fading channels with discrete input signal constellations. In particular, we have derived the optimal diversity reliability exponent for multidimensional signal constellations obtained from the rotation of classical complex-plane constellations, and we have shown that there is a tradeoff between the transmission rate, optimal achievable diversity, dimension of the rotations and size of the complex-plane signal constellation given by a modified form of the Singleton bound. Since using rotated constellations induces an increase in decoding complexity, the Singleton bound establishes the optimal rate-diversiy-complexity tradeoff. We have shown that practical coding schemes can achieve the optimal rate-diversity-complexity tradeoff.
%%%%%%%%%%%%%%%%%%%%%%%%%%%%%%%%%%%%%%%%%

\newpage

\appendices
\section{Notation} 

In this appendix we introduce the main notation that will be used throughout the proofs of the various results. We will also state without proof some of the basic results that are needed for our proofs. 
The exponential equality $\doteq$ and inequalities $\dot\geq$ and $\dot\leq$ were introduced in \cite{Tse}. We write 
\[f(z) \doteq z^d\]
to indicate that
\[\lim_{z\rightarrow \infty} \frac{\log f(z)}{\log z} =d.\]
The exponential inequalities $\dot\geq$ and $\dot\leq$ are defined similarly. 
For vectors $\xv,\yv\in\RR^n$, the notation $\xv \prec \yv$ is used to denote componentwise vector inequality, namely $x_i<y_i, i=1,\dotsc,n$. The inequalities $\succ,\preceq,\succeq$ are used similarly.
The function $\openone\{\Ec\}$ is the indicator function of the event $\Ec$, namely, $\openone\{\Ec\}=1$ when the event $\Ec$ is true, and zero otherwise. Sets are denoted with calligraphic font and the corresponding complements are denoted with a superscript $c$.
Similarly to \cite{Tse} we have the following.
\begin{definition} 
The {\em normalized} fading coefficients are defined as
\[
\alpha_b \eqdef -\frac{\log \gamma_b}{\log\snr} \;\;\;\; b=1,\dotsc,B.
\]
\end{definition}
Then, from \cite{nguyen2007} we have that
\begin{proposition}
The joint distribution of the vector $\alphav=(\alpha_1,\dotsc,\alpha_B)$ is given by
\beq
p(\alphav) = \left(\frac{m^m\log\snr}{\Gamma(m)}\right)^B e^{-m\sum_{b=1}^B \snr^{-\alpha_b}}
\snr^{-m\sum_{b=1}^B \alpha_b}
\eeq
and in the limit for large $\snr$, behaves as
\beq
p(\alphav) \doteq \snr^{-m\sum_{b=1}^B\alpha_b}
\eeq
for $\alphav\in\RR_+^B$.
\end{proposition}
\begin{definition} 
The $k$-th vector of {\em normalized} fading coefficients is defined as
\[
\alphav_k \eqdef (\alpha_{N(k-1)+1},\dotsc,\alpha_{Nk})\ \;\;\;\; k=1,\dotsc,K.
\]
\end{definition}
%%%%%%%%%%%%%%%%%%%%%%%%%%%%%%%%%%%%%%%%%
\newpage
\section{Proof of Proposition \ref{prop:exponent_singleton_lattice_ub}}
\label{appendix:proof_prop_ub}

An upper bound to the mutual information yields a lower bound on the outage probability, and thus, an upper bound to the reliability exponent. Since all rotations induce an $N\times N$ MIMO channel,  from \eqref{eq:mi_rotation} we obtain,
\begin{align}
I(\snr,\Hm) &\leq \frac{1}{K}\sum_{k=1}^K \frac{1}{N} \min\left\{N M, \log\det(\Id + \snr\;\Hm_k \Mm_k \Mm_k^\dag \Hm_k^\dag)\right\}\\
&=\frac{1}{K}\sum_{k=1}^K  \min\left\{M, \frac{1}{N}\sum_{n=1}^{N}\log(1 + \snr \gamma_{N(k-1)+n})\right\}.
\end{align}

Now, we can express the outage probability as
\begin{align}
\Pout&= \Pr(I(\snr,\Hm)< R)\\
&\geq \Pr\left(\frac{1}{K}\sum_{k=1}^K  \min\left\{M, \frac{1}{N}\sum_{n=1}^{N}\log(1 + \snr \gamma_{N(k-1)+n})\right\} < R\right)\\
&\doteq \Pr\left(\frac{1}{K}\sum_{k=1}^K  \min\left\{M, \frac{\log\snr}{N}\sum_{n=1}^{N}[1-\alpha_{N(k-1)+n}]_+\right\} < R\right)\label{eq:step3}\\
&\dot\geq \int_{\Oc_\epsilon\cap\RR_+^B} \snr^{- m\sum_{b=1}^B \alpha_b} d\alphav \label{eq:step4}
\end{align}
where \eqref{eq:step3} follows from $(1 + \snr \gamma_{N(k-1)+n}) \doteq [1-\alpha_{N(k-1)+n}]_+$, $[x]_+ = \max(0,x)$ denotes the positive part of $x\in\RR$, and
\beq
\Oc_\epsilon \eqdef \left\{\alphav\in\RR^B \;\; : \;\; \frac{1}{K}\sum_{k=1}^K\openone\{\alphav_k \succeq \onev + \epsilonv\}>1-\frac{R}{M}\right\}
\eeq
denotes the large $\snr$ outage event, and where $\onev = (1,\dotsc,1)$ and $\epsilonv=(\epsilon,\dotsc,\epsilon)$ both of dimension $N$. Note that \eqref{eq:step4} is valid for any $\epsilon>0$ and in particular for $\epsilon\to 0$.
Using Varadhan's integral lemma \cite{dembo_zeitouni}, we obtain,
\begin{align}
d_\Xc\leq d_{\rm out} &=
 - \lim_{\snr\to\infty} \frac{1}{\log\snr} \log \left( \int_{\Oc_\epsilon\cap\RR_+^B} \snr^{- m\sum_{b=1}^B \alpha_b} d\alphav\right)\\
&= - \lim_{\snr\to\infty} \frac{1}{\log\snr} \log \left( \int_{\Oc_\epsilon\cap\RR_+^B} \log\snr \exp\left(- m\sum_{b=1}^B \alpha_b\right) d\alphav\right)\\
&=\inf_{\Oc_\epsilon\cap\RR_+^B} \left\{m\sum_{b=1}^B \alpha_b\right\}
\end{align}
It is not difficult to show that $d_{\rm out}= m \,\kappa\, N$, where $\kappa$ is the unique integer such that
\beq
\kappa < K\left(1-\frac{R}{M}\right)\leq \kappa+1.
\eeq
Hence we get that
\beq
d_\Xc \leq d_{\rm out}= m\, N\,\left(1+\left\lfloor \frac{B}{N}\left(1-\frac{R}{M}\right) \right\rfloor\right)
\eeq
which is precisely the desired result.

%%%%%%%%%%%%%%%%%%%%%%%%%%%%%%%%%%%%%%%%%
\newpage
\section{Proof of Proposition \ref{prop:exponent_singleton_lattice_lb}}
\label{appendix:proof_prop_lb}

For any two codewords $\Xm(0),\Xm(1) \in \Xc$, we can write that the pairwise error probability
\begin{align}
P(\Xm(0)\to\Xm(1)|\Hm) &\leq \exp\left(-\frac{\snr}{4}\left \| \Hm(\Xm(0)-\Xm(1))  \right \|^2\right)\\
&=\prod_{k=1}^K \exp\left(-\frac{\snr}{4}\left \| \Hm_k\Mm_k(\Sm_k(0)-\Sm_k(1))\right \|^2\right)
\end{align}
where $\Sm_k(i)$ is such that the portion of codeword rotated by the $k$-th matrix is $\Xm_k(i)=\Mm_k \Sm_k(i)$, and
$\Hm = \diag(\Hm_1,\dotsc,\Hm_K)$. Assuming that the entries of $\Sm_k(0)$ and $\Sm_k(1)$ are chosen i.i.d. with uniform distribution over $\Sc$, we have that the ensemble pairwise error probability can be expressed as
\beq
\overline {P(\Xm(0)\to\Xm(1)|\Hm)} \leq \prod_{k=1}^K \left[ \frac{1}{2^{2MN}}\sum_{\sv\in\Sc^{N}} \sum_{\sv'\in\Sc^{N}} \exp\left(-\frac{\snr}{4}\left \| \Hm_k\Mm_k(\sv-\sv')\right \|^2\right) \right]^L.
\eeq

Similarly to \cite{albert_beppe_it}, summing over the $2^{LBR}-1$ codewords different from the $0$ message we have that
\begin{align}
\overline {P_e(\snr|\Hm)} &\leq 2^{LBR}\prod_{k=1}^K \left[ \frac{1}{2^{2MN}}\sum_{\sv\in\Sc^{N}} \sum_{\sv'\in\Sc^{N}} \exp\left(-\frac{\snr}{4}\left \| \Hm_k\Mm_k(\sv-\sv')\right \|^2\right) \right]^L\\
&= \exp \left(-BLM\log2 \;\;E(\snr,\alphav)    \right)
\end{align}
where the exponent $E(\snr,\alphav)$ is given by
\beq
E(\snr,\alphav) = 1 - \frac{R}{M} - \frac{1}{BM} \sum_{k=1}^K \log_2\left(1+ \frac{1}{2^{MN}}\sum_{\sv'\neq\sv} e^{-\frac{1}{4} \sum_{n=1}^{N} \snr^{1-\alpha_{N(k-1)+n}} |\tilde{x}_{k,n}|^2 }\right)
\eeq
and $\tilde{\xv}_{k} = \Mm_k(\sv-\sv') = (\tilde{x}_{k,1},\dotsc,\tilde{x}_{k,N})^T$ is the rotated difference vector. We now assume that the rotation matrices have {\em full diversity}. That implies that all the components of the rotated difference vector $\tilde{\xv}_{k}$ are different from zero. 
Then, for full diversity rotations we have that
\begin{align}
&\log_2\left(1+ \frac{1}{2^{MN}}\sum_{\sv'\neq\sv} e^{-\frac{\max_k\{|\tilde{\xv}_k|^2\}}{4} \sum_{n=1}^{N} \snr^{1-\alpha_{N(k-1)+n}}}\right)\\
&\leq\log_2\left(1+ \frac{1}{2^{MN}}\sum_{\sv'\neq\sv} e^{-\frac{1}{4} \sum_{n=1}^{N} \snr^{1-\alpha_{N(k-1)+n}} |\tilde{x}_{k,n}|^2}\right)\\
&\leq \log_2\left(1+ \frac{1}{2^{MN}}\sum_{\sv'\neq\sv} e^{-\frac{\min_k\{|\tilde{\xv}_k|^2\}}{4} \sum_{n=1}^{N} \snr^{1-\alpha_{N(k-1)+n}}}\right).
\end{align}
For large $\snr$ both bounds have the same behavior, and thus we have that
\beq
\lim_{\snr\to\infty} \log_2\left(1+ \frac{1}{2^{MN}}\sum_{\sv'\neq\sv} e^{-\frac{1}{4} \sum_{n=1}^{N} \snr^{1-\alpha_{N(k-1)+n}} |\tilde{x}_{k,n}|^2}\right) = \begin{cases}
MN \;\;\;\;\text{if $\alphav_k \succ \onev$}\\
0 \;\;\;\;\;\;\;\;\;\;\text{otherwise}
\end{cases}
\eeq
where $\alphav_k = (\alpha_{N(k-1)+1},\dotsc,\alpha_{Nk})^T$ and hence
\beq
\overline {P_e(\snr|\Hm)} \dot\leq \exp \left(-BLM\log2 \;\;E_\delta(\alphav)  \right)\\
\eeq
where
\begin{align}
E_\delta(\alphav) &\eqdef 1 - \frac{R}{M} - \frac{N}{B}\sum_{k=1}^K \openone\{\alphav_k \succeq \onev-\deltav\}= 1 - \frac{R}{M} - \frac{1}{K}\sum_{k=1}^K \openone\{\alphav_k \succeq \onev-\deltav\}
\end{align}
and $\deltav=(\delta,\dotsc,\delta)\in\RR_+^{N}$. We now define the large $\snr$ error event as
\begin{align}
\Ec_\delta &= \left\{ \alphav\in\RR^B \;\;\;:\;\;\; E_\delta(\alphav) \leq 0\right \}\\
& = \left\{ \alphav\in\RR^B \;\;\;:\;\;\; \sum_{k=1}^K \openone\{\alphav_k \succeq \onev-\deltav\} \geq K\left(1-\frac{R}{M}\right)\right \}.
\end{align}
Using the previous results we write that,
\begin{align}
\overline{P_e(\snr)} \;&\dot\leq \;\int_{\alphav\in\RR_+^B} \snr^{-m\sum_{b=1}^B \alpha_b} \min\left\{1,\exp \left(-BLM\log2 \;E_\delta(\alphav) \right)\right\} d\alphav\\
&= \int_{\alphav\in\Ec_\delta\cap\RR_+^B} \snr^{-m\sum_{b=1}^B \alpha_b} d\alphav\\
&+ \int_{\alphav\in\Ec_\delta^c\cap\RR_+^B} \snr^{-m\sum_{b=1}^B \alpha_b} \exp \left(-BLM\log2 \;E_\delta(\alphav) \right) d\alphav
\end{align}
In a similar way to the proof of Lemma \ref{lemma:gi} the probability of two randomly chosen codewords over $\Sc$ being the same is strictly greater than zero, and goes to zero only for $L\to\infty$. We now study how large $L$ has to be in order for this event not to dominate the overall error probability. If we let 
\beq
\lambda = \lim_{\snr \to \infty} \frac{L}{\log \snr}
\eeq
we can write
\begin{align}
\overline{P_e(\snr)} &\dot\leq \int_{\alphav\in\Ec_\delta\cap\RR_+^B} \snr^{-m\sum_{b=1}^B \alpha_b} d\alphav \label{eq:exp1}\\
&+ \int_{\alphav\in\Ec_\delta^c\cap\RR_+^B} \exp \left(-\log\snr\left[m\sum_{b=1}^B \alpha_b + \lambda BM\log2 \;E_\delta(\alphav) \right] \right)d\alphav
\label{eq:exp2}
\end{align}
Therefore, the overall random coding exponent is given by the minimum of the exponents of \eqref{eq:exp1} and \eqref{eq:exp2},
\beq
d_\Xc(R) \geq d_\Xc^{(r)}(R)=\sup_{\delta>0} \min \left\{d_\Xc^{(r),\infty}(R),d_\Xc^{(r),\lambda}(R)\right\}
\label{eq:sup1}
\eeq
where
\beq
d_\Xc^{(r),\infty}(R) = \inf_{\alphav\in\Ec_\delta\cap\RR_+^B} m\left\{\sum_{b=1}^B \alpha_b\right\}
\eeq
is the exponent corresponding to \eqref{eq:exp1} and
\beq
d_\Xc^{(r),\lambda}(R) = \inf_{\alphav\in\Ec_\delta^c\cap\RR_+^B} \left\{m\sum_{b=1}^B \alpha_b + \lambda BM\log2 \;E_\delta(\alphav)\right\}
\label{eq:dlambda}
\eeq
is the exponent that characterizes the effect of finite length \eqref{eq:exp2}. It is not difficult to show that the first infimum is achieved by $\kappa$ vectors $\alphav_k\succeq \onev-\deltav$, where $\kappa$ is the unique integer such that
\beq
\kappa-1 < \left\lceil K \left(1-\frac{R}{M}\right)\right\rceil \leq \kappa
\eeq
resulting in the exponent being
\beq
d_\Xc^{(r),\infty}(R) = (1-\delta)\, m\, N\, \left\lceil \frac{B}{N} \left(1-\frac{R}{M}\right)\right\rceil.
\eeq
As for the second exponent \eqref{eq:dlambda}, we can rewrite it as follows
\begin{align}
d_\Xc^{(r),\lambda}(R) &=  \lambda BM\log2 \left(1-\frac{R}{M}\right)\\
& + \inf_{\alphav\in\Ec_\delta^c\cap\RR_+^B} \left\{m\sum_{b=1}^B \alpha_b - \lambda BM\log2\, \frac{1}{K}\sum_{k=1}^K \openone\{\alphav_k \succeq \onev-\deltav\} \right\}\\
&=  \lambda BM\log2 \left(1-\frac{R}{M}\right)\\
& + m \,\inf_{\alphav\in\Ec_\delta^c\cap\RR_+^B} \left\{\sum_{k=1}^K \left( \sum_{n=1}^N \alpha_{k,n} - \frac{\lambda NM\log2}{m} \, \openone\{\alphav_k \succeq \onev-\deltav\} \right) \right\}
\end{align}
The constraint set $\Ec_\delta^c$ is defined as follows
\beq
\Ec_\delta^c \eqdef \left\{ \alphav\in\RR^B \;\;\;:\;\;\; \sum_{k=1}^K \openone\{\alphav_k \succeq \onev-\deltav\} < K\left(1-\frac{R}{M}\right)\right \}.
\eeq
We distinguish two cases. When $0\leq \lambda NM\log2<m$ then the terms
\beq
\sum_{n=1}^N \alpha_{k,n} - \frac{\lambda NM\log2}{m} \, \openone\{\alphav_k \succeq \onev-\deltav\}
\eeq
attain its minimum value for $\alphav_k=\zerov$. On the other hand, when $\lambda NM\log2 \geq m$, the constraint set dictates that there should be 
\beq
\kappa = \left\lfloor K\left(1-\frac{R}{M}\right)\right\rfloor
\eeq
vectors $\alphav_k\succeq \onev-\deltav$, and the infimum becomes
\beq
\lambda BM\log2 \left(1-\frac{R}{M}\right) + m \left\lfloor K\left(1-\frac{R}{M}\right)\right\rfloor \left(N(1-\delta)-\frac{\lambda NM\log2}{m}  \right).
\eeq
Combining the previous results and noting that the supremum in \eqref{eq:sup1} is achieved for $\delta\to 0$, we find the desired result.
%%%%%%%%%%%%%%%%%%%%%%%%%%%%%%%%%%%%%%%%%

\newpage

\newpage

\newpage

\newpage

\newpage

\newpage

\newpage

\newpage

%%%%%%%%%%%%%%%%%%%%%%%%%%%%%%%%%%%%%%%%%

\newpage
\bibliographystyle{IEEE}

%%%%%%%%%%%%%%%%%%%%%%%%%%%%%%%%%%%%%%%%%

\end{document}